# Network Clustering for Multi-task Learning

Dehong Gao, Wenjing Yang, Huiling Zhou, Yi Wei, Yi Hu and Hao Wang
Alibaba Group
Hangzhou City, Zhejiang Province, China
{dehong.gdh, carrie.ywj, zhule.zhl, yi.weiy, erwin.huy, longran.wh}@alibaba-inc.com

## ABSTRACT

The Multi-Task Learning (MTL) technique has been widely studied by word-wide researchers. The majority of current MTL studies adopt the hard parameter sharing structure, where hard layers tend to learn general representations over all tasks and specific layers are prone to learn specific representations for each task. Since the specific layers directly follow the hard layers, the MTL model needs to estimate this direct change (from general to specific) as well. To alleviate this problem, we introduce the novel cluster layer, which groups tasks into clusters during training procedures. In a cluster layer, the tasks in the same cluster are further required to share the same network. By this way, the cluster layer produces the general presentation for the same cluster, while produces relatively specific presentations for different clusters. As transitions the cluster layers are used between the hard layers and the specific layers. The MTL model thus learns general representations to specific representations gradually. We evaluate our model with MTL document classification and the results demonstrate the cluster layer is quite efficient in MTL.

## 1 Introduction

Different from the single task learning, Multi-Task Learning (MTL) trains a united model over several tasks simultaneously [4]. Recently, it has been noticed that MTL particularly with Deep Learning (DL) has led to many successes in Natural Language Processing (NLP) and Computer Vision (CV) [11, 18].

There mainly exist two network structures in current DL-based MTL approaches, i.e., the soft structure and the hard structure as shown in Figure 1 [15]. The soft structure requires that each task has its own parameters. In soft layers, the parameters of different tasks are regularized in order to learn relationships between tasks. In specific layers, the parameters are specified to each task. The hard structure forces the MTL model to firstly learn a unit representation over all tasks. Specific layers follow hard layers for each task as well. Compared with the soft structure, the hard structure significantly reduces the model size and the risk of overfitting. Thus, the hard structure is more popular in industry [3]. In the hard structure, the hard layers are expected to learn general representations over all tasks, while the specific layers are expected to learn specific representations for each task. Since the specific layers directly follow with the hard layers, the MTL model needs to first estimate the general representations and then estimate the specific representations alternatively. To certain degree, this rapid change is difficult for the MTL model to estimate without any transition.

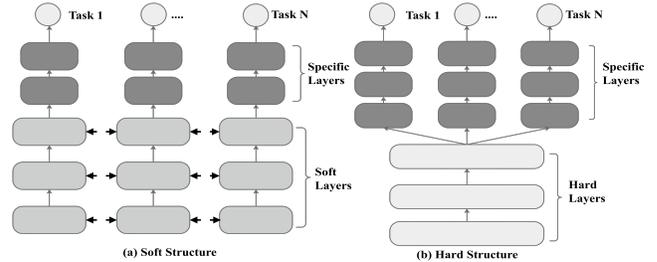

**Figure 1: The soft and hard parameter sharing structures**

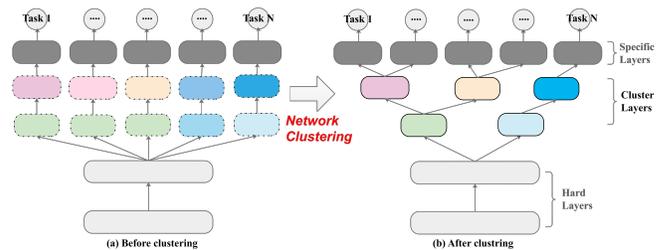

**Figure 2: The proposed cluster parameter sharing structure**

To alleviate this problem, we propose a novel parameter sharing mechanism in this paper. To certain degree, the tasks that are related tend to learn similar parameters in the MTL model [15]. Inspired by this idea, we present a novel neural layer, i.e., the cluster layer. As shown in Figure 2.(a), the MTL model first estimates the specific parameters for each task and the similar tasks tend to learn the similar parameters (with similar colors). Then, we perform network clustering to group tasks with similar parameters into clusters. For tasks in one cluster, their estimated parameters are further replaced with the same parameters (i.e., the parameter center of all tasks in this center) as seen in Figure 2.(b). There are three main advantages of this cluster sharing mechanism. First, like hard layers, the cluster layer learns general representations for all tasks intra one cluster. Like specific layers, the cluster layer learns specific representations for inter clusters. The cluster layer is regarded as a transition from the hard layers to the specific layers. The MTL model thus learns representations from general to specific gradually. Second, the cluster layer significantly reduces the parameter size. The parameter size of a specific layer for MTL is $N$ times larger than that for a single task ($N$ is the task number). Replacing with the cluster layer, the parameter size is reduced to $K/N$ times less than that of a specific layer ($K$ is the cluster number). This will benefit a lot to massive multi-task learning (MMTL) [14] (where MTL confronts the learning problem over tens of or hundreds of tasks e.g., Drug discovery [13]) since the parameters will scale up rapidly as the increasing of task number. Finally, this cluster sharing mechanism is regarded as a regularization in the MTL model, which is good for the MTL model to reduce the over-fitting risk. In this paper, we also present the learning algorithm of the Network Clustering MTL model, which alternatively performs model estimation and network clustering during train procedures.

Experiments are conducted on multi-task document classification with public datasets. The results show the cluster layer is quite efficient in MTL. The main contributions of this paper are summarized as follows: 1) As far as we know, we are the first to cluster the cluster layer for MTL. By this way, the MTL model gradually learns general representations to specific representations. 2) We present the NCMTL approach, which alternatively performs network clustering and parameter estimation.

## 2 Network Clustering MTL

The task of multi-task learning is mathematically formulated as: Given a serial of tasks $T = \{T_1, T_2, ..., T_N\}$, we aim to learn a united mapping function $Y = f(X, \theta)$, which outputs labels $Y_i$ with the inputs $X_i$ of the task $T_i$. In our Network Clustering MTL (NCMTL) model, $\theta$ is written as $\{W_H, W_c, W_s\}$, which denote the parameters in the hard, cluster and specific neural networks respectively. Meanwhile, the cluster layers are set between the hard layers and the specific layers as shown in Figure 2.(b).

For the hard layers $W_H$, it is possible to adopt the recurrent and/or convolutional neural networks as the backbone networks. But the hard layers are not the key interest of this paper. With the limited space, we focus more on the cluster layers.

For the cluster layers $W_c$, we incorporate $L$ cluster layers in the NCMTL model. Taking the cluster layer $l_i$ as an example, we first assume there exist *specific weights* $w'_{ij}$ for the task $T_j$. Since similar tasks tend to learn similar parameters, all tasks are grouped into $K_i$ clusters with the similarities of their specific weights $w'_{i*}$ (* denotes all tasks, and $K_i$ is the predefined cluster number in this cluster layer). We then obtain the parameter center of each cluster. Let $c_{ik}$ be the parameter center of the $k^{th}$ cluster, the original specific weights of the tasks in this cluster are replaced with $c_{ik}$. By this way, the tasks in one cluster share the same parameters. Compared with the hard layer, which produce the same representation for all tasks, the cluster layer produces the same representation for a group of tasks. When we gradually increase the cluster numbers of the cluster layers, the NCMTL model is prone to learn the general features to the specific features.

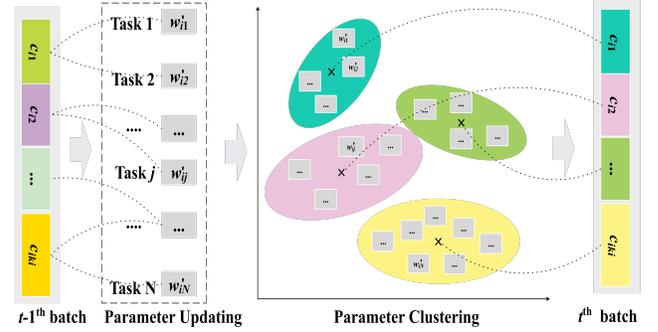

**Figure 3: Parameter updating and clustering**

To save the memory usage, the final NCMTL model does not need to include the specific weights $w'_{i*}$. We can only keep the parameter centers and the cluster results in the final NCMTL model. Since we use the $K_i$ parameter centers replacing the specific weights of all $N$ tasks, the parameter size is reduced to $K_i/N$ in the cluster layer $l_i$ less than that in the specific layer.

In practice, the clustering procedure is performed during training stages. As shown in Figure 3, at the $(t-1)^{th}$ training batch we update $w'_{ij}$ with the parameter center $c_{ik}$ of task $j$. Then clustering is carried out on the updated $w'_{i*}$ and forms the new cluster center $c_{ik}$ of the next $t^{th}$ training batch. Under the hard cluster model (i.e., one task is grouped into only one cluster in a cluster layer), the clustering loss function is presented by

$$\mathcal{L}_c(\theta) = \sum_{i=1}^{L} \sum_{j=1}^{N} \sum_{k=1}^{k_i} r_{ijk} \|w'_{ji} - c_{ki}\|^2$$

$$r_{ijk} = \begin{cases} 1 & if \ T_j \in c_{ik} \\ 0 & else \end{cases}$$

where $r_{ijk}$ are the cluster results, denoting that the task $T_j$ is grouped into the cluster $c_{ik}$ of the cluster layer $l_i$.

For the specific layers $W_s$, it predicts the final probabilities $Y_j$ given the input $X_j$ of the task $T_j$. This is a standard classification problem. With the expected labels $\hat{Y}_j$, the classification loss function is defined by

$$\mathcal{L}_p(\theta) = -\sum_{j=1}^{N} \hat{Y}_j \log (Y_j)$$

In sum, the loss function of our NCMTL model is written as

$$\mathcal{L}(\theta) = \mathcal{L}_p(\theta) + \alpha \mathcal{L}_c(\theta)$$

where $\alpha$ balances the classification loss and the clustering loss. To some extent, $\mathcal{L}_c(\theta)$ acts as the regularization to the NCMTL model. We alternatively perform the model estimation and network clustering as shown in Algorithm 1.

## 3. Experiments

### 3.1 Datasets

In this paper, our NCMTL model is evaluated with MTL document classification. 14 Amazon product reviews datasets[1] are selected from the domains of {"apparel", "baby", "books", "camera", "DVD", "electronic", "health", "kitchen", "magazines", "music", "software", "sports", "toys" and "video"}[2]. These datasets are document-level reviews, and the goal is to predict the sentiment (i.e., positive or negative) of each document. We totally collect 27,755 documents for all 14 domains. Each dataset contains 37K vocabularies on average, and each document includes 115 words averagely. For each dataset, we equally split it into ten sets, among which nine are used as the training set and the rest is used as the testing set. Accuracy is employed to evaluate the performances of our NCMTL model.

Our proposed approach is compared with the baseline approaches, including the single-task learning based and the multi-task learning based.

- LSTM: The standard LSTM-based neural network for single task document classification [6].
- TextCNN: The standard convolution neural network for single task document classification [7].
- HAN: The hierarchical sentence-level and document-level attention network for single task document classification [17].
- SoftShare: The soft parameter sharing approach for MTL document classification [16]. For each task, we adopt the network structure similar to the HAN approach in the soft layers. The soft layers of different tasks are regularized with $\ell_2$ normalization. The specific layers are used for classification of each task.
- HardShare: The hard parameter sharing approach for MTL document classification [12]. In the hard layers, we adopt the same network configure like the soft layers in the SoftShare approach. The specific layers in the HardShare approach are same to those of the SoftShare approach.
- ASP-MTL: The hard parameter sharing approach for MTL document classification [10]. An adversarial task is introduced to alleviate the feature interaction between tasks.
- NCMTL: Our cluster layer-based MTL. Different to the HardShare approach, network clustering is performed on all specific layers except the last one. KMeans++ [1] is adopted in network clustering of each training batch.

**Implementation detail**: For a fair comparison, we reimplement all the above approaches with nearly the same hyperparameter settings. The word embedings are initialized with pre-trained GloVe vectors like [11]. The dimension of the word embedding is 200. In these experiments, we focus on the comparison of the cluster layers, rather than the comparison of manual-designed &

deliciated network structure. Thus, we adopted relative-simple network structure as base model. Two word-level and two sentence-level LSTM layers are adopted as the hard layers. We adopt three cluster layers with the hidden sizes {32, 32, 16} for each task and the cluster numbers {3, 5, 10}. Finally, one specific layer is used for the perdition of each task. The approaches are trained on Telsa V100 GPU with Adam optimizer (learning rate $1E - 5$). The batch size is set to 32.

---

**Algorithm 1** Network Clustering MTL

**Input**: the loss functions $\mathcal{L}(\theta)$, and training datasets of all tasks

**Parameter**: the model parameters $\theta = \{W_H, W_c, W_s\}$

**Output**: $\theta$

1: **For** each batch of train data **do**
2:     Feed the train batch into the NCMTL model to get $\mathcal{L}$
3:     Update the hard and specific parameters $W_H$ and $W_S$ with SGD, e.g., $W_H = W_H - \eta\, \partial \mathcal{L}_p / \partial W_H$
4:     Update the specific parameter $W_c'$ of the cluster layers with SGD, i.e., $W_c' = W_c' - \eta\, \partial \mathcal{L}_c / \partial W_c$
5:     **For** each cluster layer $l_i$ **do**
6:         Cluster tasks on the specific weights $w_{i*}'$ and get $r_{i*}$
7:         Compute parameter centers $c_{i*}$ and update $w_{i*}'$ by $r_{i*}$
8:     **End For**
9: **End For**
10: **Return** model parameter $\theta$

---

### 3.2 Analysis of Results

All approaches run three times and we show the average accuracy in Table 1. Compared with the baseline approaches, the NCMTL approach achieves the significant improvement except the ASP-MTL approach. Thus, the experiment results illustrate the efficiency of the cluster layer.

We show the cluster results of some cluster layers in our final NCMTL model in Figure 4. To certain degree, the cluster results verify our previous hypotheses that the similar tasks prone to learn similar parameters. For example, "camera", "electronic" and "software" are related to the electronic category and these tasks are in the same cluster. Meanwhile, we found the clusters in the cluster layer 1 are more general than those in the cluster layer 2, e.g., the cluster {"apparel", "baby", "books", "magazines"} in the cluster layer 1 compared with the clusters {"apparel", "baby"} and {"books", "magazines"} in the cluster layer 2. To certain degree, this verifies our motivation that the cluster layers gradually learn the general representations to the specific representations. We also trace the change of the clustering results during the training iterations. It is found the clustering results are relatively stable, which means certain tasks (e.g., "music" and "video" classification tasks) are always grouped to the same cluster after a few training iterations. To accelerate the training

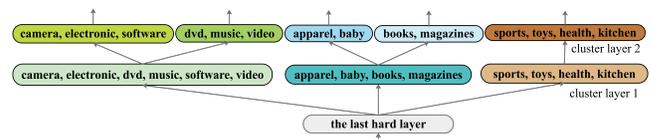

**Figure 4: Some cluster results in our final NCMTL model**

| Datasets | Single Task | | | | Multiple Tasks | | | |
|---|---|---|---|---|---|---|---|---|
| | LSTM | TextCNN | HAN | Single Avg | SoftShare | HardShare | ASP-MTL | NCMTL |
| Apparel | 84.0 | 83.9 | 84.9 | 84.3 | 85.2 | 86.1 | 87.1 | 86.6 |
| Baby | 84.5 | 85.8 | 86.9 | 85.7 | 88.7 | 88.3 | 88.2 | 88.8 |
| Books | 78.4 | 84.3 | 85.8 | 82.8 | 86.5 | 86.9 | 86.2 | 86.7 |
| Camera | 84.7 | 87.1 | 86.8 | 86.8 | 88.4 | 87.7 | 88.2 | 89.1 |
| DVD | 81.4 | 81.7 | 83.3 | 82.1 | 85.0 | 85.8 | 86.5 | 88.5 |
| Electronic | 80.0 | 80.2 | 80.9 | 80.4 | 84.2 | 86.1 | 85.8 | 86.3 |
| Health | 83.5 | 84.4 | 85.7 | 84.5 | 86.1 | 86.7 | 88.2 | 87.0 |
| Kitchen | 78.4 | 84.3 | 85.8 | 82.8 | 86.2 | 85.9 | 86.2 | 87.1 |
| Magazines | 88.7 | 88.9 | 89.4 | 89.0 | 90.0 | 89.3 | 90.2 | 90.6 |
| Music | 76.5 | 81.2 | 82.5 | 80.1 | 82.8 | 82.6 | 84.5 | 84.2 |
| Software | 83.9 | 84.9 | 85.7 | 84.8 | 85.6 | 87.8 | 87.2 | 88.4 |
| Sports | 80.9 | 83.2 | 84.2 | 82.8 | 85.1 | 86.4 | 86.7 | 87.3 |
| Toys | 82.9 | 84.3 | 85.3 | 84.2 | 85.4 | 86.6 | 88.0 | 87.5 |
| Video | 80.9 | 84.3 | 85.4 | 83.5 | 86.8 | 87.3 | 87.5 | 88.3 |
| AVG | 82.1 | 84.2 | 85.3 | 84.0 | 86.1 | 86.6 | 87.2 | **87.6** |

Table 1: Accuracy of the NCMTL model on 14 Datasets against Baselines

speed, we freeze the cluster results after four training epochs.

## 4. Related Work

Multi-Task Learning (MTL) aims to train a united model over several tasks simultaneously [4]. In NLP and CV, MTL especially with Deep Learning has led to many successes recently [11, 18]. In this paper, we thus focus on the DL-based MTL approaches.

Ruder summarizes the DL-based MTL network structures, which divides into two categories, i.e., the soft parameter sharing structure and the hard parameter sharing structure [15]. For example, the soft structure is selected in [16] to regularize the parameters of all models, while the hard structure is adopted in [12] to share the convolution backbone network over all tasks. Specially for MTL text classification, two mechanisms (i.e., an external memory and a reading/writing communication) are introduced in to share the task information [8, 9]. Liu et al. employ the adversarial mechanism to overcome the task differences in [10]. The meta-knowledge from different tasks is extracted and used in the MTL learning in [5]. There are many studies on MTL document classification [11]. With limited spaces, we only focus the MTL network structure. These methods mainly adopt the hard structure, of which the hard and specific layers tend to learn general and specific representations, respectively. In this paper, we present the cluster layers, which is regarded as a transition from the hard layers to the specific layers for MTL and learn general representations to specific representations.

## 5. Conclusion

In this paper, we propose the novel cluster layer to address the MTL problems. The cluster layer automatically groups the network parameters of similar tasks. The cluster layer not only reduces the model size, but also learns representations from general ones to specific ones gradually. The experimental results show that the proposed NCMTL is quite efficient in multi-task document classification. The cluster results verify that the cluster layers learn general representations to specific representations.

Currently network clustering is performed within layers. We will attempt to cluster networks across layers. Besides, network clustering can be regarded as a method of automatic Network Architecture Search (NAS) [19]. We are exploring the opportunities to extend our NCMTL with the NAS approaches.

# Insert Your Title Here